\documentclass[11pt]{article}
\usepackage{a4p}
\usepackage{amssymb}
\usepackage{epsfig}
\usepackage{cite}
\usepackage{pennames}
\usepackage{rotating}
\newcommand{\intLdtfull}  {57.21}
\newcommand{\dLstat}  {0.15}
\newcommand{\dLsys}   {0.20}
\newcommand{\rroots}  {182.68}
\newcommand{\GENTfull} {15.71}
\newcommand{\epem}{\ensuremath{\mathrm{e}^+\mathrm{e}^-}}

\newcommand{\Zz}{\ensuremath{{\mathrm{Z}^0}}}

\newcommand{\WW}{\ensuremath{\mathrm{W}^+\mathrm{W}^-}}
\newcommand{\Wp}{\ensuremath{\mathrm{W}^+}}

\newcommand{\Wm}{\ensuremath{\mathrm{W}^-}}
\newcommand{\eeWW}{\ensuremath{\epem\rightarrow\WW}}
\newcommand{\qq}{\ensuremath{\mathrm{q\overline{q}^\prime}}}

\newcommand{\lnu}{\ensuremath{\ell\overline{\nu}_{\ell}}}
\newcommand{\lpnu}{\ensuremath{\ell^+ \nu_{\ell}}}
\newcommand{\lmnu}{\ensuremath{{\ell^{\prime}}^-\overline{\nu}_{\ell^{\prime}}}}
\newcommand{\enu}{\ensuremath{\mathrm{e\overline{\nu}_{e}}}}

\newcommand{\qqqq}{\ensuremath{\qq\qq}}
\newcommand{\qqln}{\ensuremath{\qq\lnu}}
\newcommand{\Ks}    {\ensuremath{\mathrm{K}^0_{\mathrm{s}}}}
\newcommand{\WWqqln}{\ensuremath{\WW\rightarrow\qq\lnu}}
\newcommand{\WWqqqq}{\ensuremath{\WW\rightarrow\qq\qq}}

\newcommand{\WWlnln}{\ensuremath{\WW\rightarrow\lpnu\lmnu}}

\newcommand{\Mw}{\ensuremath{M_{\mathrm{W}}}}
\newcommand{\Gw}{\ensuremath{\Gamma_{\mathrm{W}}}}

\newcommand{\JETSET}{\mbox{J{\sc etset}}}
\newcommand{\VNI}{\mbox{V{\sc ni}}}

\newcommand{\KORALW}{\mbox{K{\sc oralw}}}

\newcommand{\EXCALIBUR}{\mbox{E{\sc xcalibur}}}
\newcommand{\grcff}{\mbox{grc4f}}
\newcommand{\Pythia}{\mbox{P{\sc ythia}}}
\newcommand{\PYTHIA}{\mbox{P{\sc ythia}}}
\newcommand{\SK}{\mbox{Sj\"ostrand-Khoze}}
\newcommand{\SKI}{\mbox{SK I}}
\newcommand{\SKII}{\mbox{SK II}}
\newcommand{\SKIIpr}{\mbox{SK II$^{\prime}$}}
\newcommand{\ARMII}{\mbox{AR 2}}
\newcommand{\ARMIII}{\mbox{AR 3}}
\newcommand{\GH}{\mbox{Gustafson and H\"akkinen}}
\newcommand{\EG}{\mbox{Ellis-Geiger}}
\newcommand{\Ariadne}{\mbox{A{\sc riadne}}}
\newcommand{\ARIADNE}{\mbox{A{\sc riadne}}}

\newcommand{\GENTLE}{\mbox{G{\sc entle}}}

\newcommand{\PHOJET}{\mbox{P{\sc hojet}}}

\newcommand{\Herwig}{\mbox{H{\sc erwig}}}
\newcommand{\HERWIG}{\mbox{H{\sc erwig}}}

\newcommand{\GeV}{\ensuremath{\mathrm{GeV}}}

\newcommand{\Prec}{\ensuremath{P^{\mathrm{reco}}}}

\newcommand{\Mfit}{\ensuremath{M_{\mathrm{fit}}}}
\newcommand{\Mrec}{\ensuremath{m_{\mathrm{rec}}}}
\newcommand{\Mtrue}{\ensuremath{m_{\mathrm{true}}}}

\newcommand{\roots}{\ensuremath{\sqrt{s}}}

\newcommand{\Ebeam}{\ensuremath{E_{\mathrm{beam}}}}

\newcommand{\Zgamma}{\ensuremath{\Zz/\gamma}}

\newcommand{\CC}{\mbox{{\sc CC03}}}

\def\etal{\mbox{{\it et al.}}}

\def\gappeq{\ensuremath{\mathrel{ \rlap{\raise.5ex\hbox{>}}
                      {\lower.5ex\hbox{\sim}}}}}
\newcommand{\lappeq}{\ensuremath{\mathrel{\rlap{\raise.5ex\hbox{<}}{\lower.5ex\hbox{\sim}}}}}

\newcommand{\PLB}[3]  {Phys.\ Lett.\ \textbf{B#1} (#2) #3}
\newcommand{\ZPC}[3]  {Z.\ Phys.\ \textbf{C#1} (#2) #3}
\newcommand{\EPC}[3]  {Eur.\ Phys.\ J.\ \textbf{C#1} (#2) #3}
\newcommand{\NIMA}[3] {Nucl.\ Instr.\ Meth.\ \textbf{A#1} (#2) #3}

\newcommand{\EP}[1]  {CERN-EP/{#1}}

\newcommand{\PRL}[3]  {Phys.\ Rev.\ Lett.\ \textbf{#1} (#2) #3}
\newcommand{\PRD}[3]  {Phys.\ Rev.\ \textbf{D#1} (#2) #3}
\newcommand{\NPB}[3]  {Nucl.\ Phys.\ \textbf{B#1} (#2) #3}

\newcommand{\CPC}[3]  {Comput.\ Phys.\ Commun.\ \textbf{#1} (#2) #3}
\newcommand{\Zzero}{\mbox{${\mathrm{Z}^0}$}}
\newcommand{\ZZ}{\mbox{\Zzero\Zzero}}

\def\opalabbiendi{OPAL Collaboration, G.\ Abbiendi \etal}
\def\opalackerstaff{OPAL Collaboration, K.\ Ackerstaff \etal}
\def\opalalexander{OPAL Collaboration, G.\ Alexander \etal}

\newcommand{\QQQQ}{\ensuremath{\mathrm{4q}}}
\newcommand{\QQLV}{\ensuremath{\mathrm{qq}\ell\nu}}
\newcommand{\nch}{\ensuremath{n_{\mathrm{ch}}}}
\newcommand{\nchave}{\ensuremath{\langle n_{\mathrm{ch}}\rangle}}
\newcommand{\nchQQQQ}{\ensuremath{\langle n_{\mathrm{ch}}^{\QQQQ}\rangle}}
\newcommand{\nchQQLV}{\ensuremath{\langle n_{\mathrm{ch}}^{\QQLV}\rangle}}
\newcommand{\Dnch}{\ensuremath{\Delta\langle n_{\mathrm{ch}}\rangle}}
\newcommand{\xpQQQQ}{\ensuremath{\langle x_{p}^{\QQQQ}\rangle}}
\newcommand{\xpQQLV}{\ensuremath{\langle x_{p}^{\QQLV}\rangle}}
\newcommand{\xp}{\ensuremath{x_{p}}}
\newcommand{\xpave}{\ensuremath{\langle\xp\rangle}}
\newcommand{\thrQQQQ}{\ensuremath{\langle 1-T^{\QQQQ} \rangle}}
\newcommand{\yQQQQ}{\ensuremath{\langle |y^{\QQQQ}| \rangle}}
\newcommand{\y}{\ensuremath{|y|}}
\newcommand{\Dxp}{\ensuremath{\Delta \langle x_{p} \rangle}}

\newcommand{\DQQQQ}{\ensuremath{\mathrm{D^{\QQQQ}}}}
\newcommand{\DQQLV}{\ensuremath{\mathrm{D^{\QQLV}}}}
\newcommand{\DD}{\ensuremath{\Delta\mathrm{D}}}
\begin{document}
\begin{titlepage}
 \begin{center}{\Large    EUROPEAN LABORATORY FOR PARTICLE PHYSICS
 }\end{center}\bigskip
\begin{flushright}
 CERN-PPE/98-zzz \\ December xx, 1998 \\
\end{flushright}
\bigskip\bigskip\bigskip\bigskip\bigskip
\begin{center}
 {\huge\bf \boldmath Colour reconnection studies in \eeWW\ \\
   \vspace{1mm}
    at $\roots=183$~GeV}\\
\end{center}
\bigskip\bigskip
\begin{center}{\LARGE The OPAL Collaboration
}\end{center}\bigskip\bigskip
\bigskip\begin{center}{\large Abstract}\end{center}
\noindent
 The predicted effects of final state interactions such as colour reconnection
 are investigated by measuring properties of hadronic decays of W bosons,
 recorded at a centre-of-mass energy of $\roots\simeq182.7$~GeV in the OPAL
 detector at LEP.  Dependence on the modelling of hadronic W decays is avoided
 by comparing \WWqqqq\ events with the non-leptonic component of \WWqqln\
 events.  The scaled momentum distribution, its mean value, \xpave, and that of
 the charged particle multiplicity, \nchave, are measured and found to be
 consistent in the two channels. The measured differences are:
\begin{eqnarray*}
  \Dnch = \nchQQQQ-2\nchQQLV  & = & +0.7 \pm 0.8 \pm 0.6 \\
  \Dxp  = \xpQQQQ-\xpQQLV     & = & (-0.09 \pm 0.09  \pm 0.05)\times 10^{-2}.
\end{eqnarray*} 
In addition, measurements of rapidity and thrust are performed for \WWqqqq\
events. The data are described well by standard QCD models and disfavour one
model of colour reconnection within the \Ariadne\ program.  The current
implementation of the \EG\ model of colour reconnection is excluded. At the
current level of statistical precision no evidence for colour reconnection
effects was found in the observables studied.  The predicted effect of colour
reconnection on OPAL measurements of \Mw\ is also quantified in the context of
models studied.

\bigskip\bigskip\bigskip\bigskip
\bigskip\bigskip
\begin{center}{\large
(Submitted to Physics Letters \textbf{B})
}\end{center}
\end{titlepage}

\begin{center}{\Large        The OPAL Collaboration
}\end{center}\bigskip
\begin{center}{
G.\thinspace Abbiendi$^{  2}$,
K.\thinspace Ackerstaff$^{  8}$,
G.\thinspace Alexander$^{ 23}$,
J.\thinspace Allison$^{ 16}$,
N.\thinspace Altekamp$^{  5}$,
K.J.\thinspace Anderson$^{  9}$,
S.\thinspace Anderson$^{ 12}$,
S.\thinspace Arcelli$^{ 17}$,
S.\thinspace Asai$^{ 24}$,
S.F.\thinspace Ashby$^{  1}$
% ...
}\end{center}\bigskip
\bigskip
$^{  1}$School of Physics and Astronomy, University of Birmingham,
Birmingham B15 2TT, UK
\newline
$^{  2}$Dipartimento di Fisica dell' Universit\`a di Bologna and INFN,
I-40126 Bologna, Italy
\newline
$^{  3}$Physikalisches Institut, Universit\"at Bonn,
D-53115 Bonn, Germany
\newline
$^{  4}$Department of Physics, University of California,
Riverside CA 92521, USA
\newline
$^{  5}$Cavendish Laboratory, Cambridge CB3 0HE, UK
\newline
$^{  6}$Ottawa-Carleton Institute for Physics,
Department of Physics, Carleton University,
Ottawa, Ontario K1S 5B6, Canada
\newline
$^{  7}$Centre for Research in Particle Physics,
Carleton University, Ottawa, Ontario K1S 5B6, Canada
\newline
$^{  8}$CERN, European Organisation for Particle Physics,
CH-1211 Geneva 23, Switzerland
\newline
$^{  9}$Enrico Fermi Institute and Department of Physics,
University of Chicago, Chicago IL 60637, USA
\newline
$^{ 10}$Fakult\"at f\"ur Physik, Albert Ludwigs Universit\"at,
D-79104 Freiburg, Germany
\newline
$^{ 11}$Physikalisches Institut, Universit\"at
Heidelberg, D-69120 Heidelberg, Germany
\newline
$^{ 12}$Indiana University, Department of Physics,
Swain Hall West 117, Bloomington IN 47405, USA
\newline
$^{ 13}$Queen Mary and Westfield College, University of London,
London E1 4NS, UK
\newline
$^{ 14}$Technische Hochschule Aachen, III Physikalisches Institut,
Sommerfeldstrasse 26-28, D-52056 Aachen, Germany
\newline
$^{ 15}$University College London, London WC1E 6BT, UK
\newline
$^{ 16}$Department of Physics, Schuster Laboratory, The University,
Manchester M13 9PL, UK
\newline
$^{ 17}$Department of Physics, University of Maryland,
College Park, MD 20742, USA
\newline
$^{ 18}$Laboratoire de Physique Nucl\'eaire, Universit\'e de Montr\'eal,
Montr\'eal, Quebec H3C 3J7, Canada
\newline
$^{ 19}$University of Oregon, Department of Physics, Eugene
OR 97403, USA
\newline
$^{ 20}$CLRC Rutherford Appleton Laboratory, Chilton,
Didcot, Oxfordshire OX11 0QX, UK
\newline
$^{ 22}$Department of Physics, Technion-Israel Institute of
Technology, Haifa 32000, Israel
\newline
$^{ 23}$Department of Physics and Astronomy, Tel Aviv University,
Tel Aviv 69978, Israel
\newline
$^{ 24}$International Centre for Elementary Particle Physics and
Department of Physics, University of Tokyo, Tokyo 113-0033, and
Kobe University, Kobe 657-8501, Japan
\newline
$^{ 25}$Institute of Physical and Environmental Sciences,
Brunel University, Uxbridge, Middlesex UB8 3PH, UK
\newline
$^{ 26}$Particle Physics Department, Weizmann Institute of Science,
Rehovot 76100, Israel
\newline
$^{ 27}$Universit\"at Hamburg/DESY, II Institut f\"ur Experimental
Physik, Notkestrasse 85, D-22607 Hamburg, Germany
\newline
$^{ 28}$University of Victoria, Department of Physics, P O Box 3055,
Victoria BC V8W 3P6, Canada
\newline
$^{ 29}$University of British Columbia, Department of Physics,
Vancouver BC V6T 1Z1, Canada
\newline
$^{ 30}$University of Alberta,  Department of Physics,
Edmonton AB T6G 2J1, Canada
\newline
$^{ 31}$Duke University, Dept of Physics,
Durham, NC 27708-0305, USA
\newline
$^{ 32}$Research Institute for Particle and Nuclear Physics,
H-1525 Budapest, P O  Box 49, Hungary
\newline
$^{ 33}$Institute of Nuclear Research,
H-4001 Debrecen, P O  Box 51, Hungary
\newline
$^{ 34}$Ludwigs-Maximilians-Universit\"at M\"unchen,
Sektion Physik, Am Coulombwall 1, D-85748 Garching, Germany
\newline
\bigskip\newline
$^{  a}$ and at TRIUMF, Vancouver, Canada V6T 2A3
\newline
$^{  b}$ and Royal Society University Research Fellow
\newline
$^{  c}$ and Institute of Nuclear Research, Debrecen, Hungary
\newline
$^{  d}$ and Department of Experimental Physics, Lajos Kossuth
University, Debrecen, Hungary
\newline
$^{  e}$ on leave of absence from the University of Freiburg

\newpage


\section{Introduction}
\label{Wprop}
 
 Hadronic data in \epem\ collisions can be characterised by event
 shape distributions and inclusive observables such as charged
 particle multiplicities and momentum spectra.  In addition to tests
 of Monte Carlo models, measurement of the properties of the hadronic
 sector of \WW\ decays allows the question of ``colour reconnection''
 \cite{bib:GPZ} to be addressed experimentally. The decay products of
 the two W decays may have a significant space-time overlap as the
 separation of their decay vertices at LEP2 energies is small compared
 to characteristic hadronic distance scales. In the fully hadronic
 channel this may lead to new types of final state
 interactions. Colour reconnection is the general name applied to the
 case where such final state interactions lead to colour flow between
 the decay products of the two W bosons.  At present there is general
 consensus that observable effects of such interactions during the
 perturbative phase are expected to be small \cite{bib:SK}.  In
 contrast, significant interference in the hadronisation process is
 considered to be a real possibility.  With the current knowledge of
 non-perturbative QCD, such interference can be estimated only in the
 context of specific models
 \cite{bib:GPZ,bib:PYTHIA,bib:SK,bib:GH,bib:LL,bib:ARIADNE,bib:HERWIG,bib:EG1,
 bib:VNI}. One should be aware that other final state effects such as
 Bose-Einstein correlations between identical bosons may also
 influence the observed event properties.

 It has been suggested \cite{bib:SK,bib:GH} that simple observable
 quantities, such as the charged multiplicity in restricted rapidity
 intervals, may be particularly sensitive to the effects of colour
 reconnection.  Later studies \cite{bib:HR,bib:WWOXFORD} showed that
 the initial estimates of \cite{bib:GH} were incorrect for a variety
 of reasons\footnote{The dominant effect was to have neglected the
 polarisation of the W.}.  More recently, studies using the \EG\ model
 \cite{bib:EG1,bib:VNI} suggested that the effect on the inclusive
 charged multiplicity itself may be larger than previously considered
 and that the mean hadronic multiplicity in \WWqqqq\ events, \nchQQQQ,
 may be as much as 10\% smaller than twice the hadronic multiplicity
 in \WWqqln\ events, \nchQQLV\ \cite{bib:EG2}. The visible effects of
 such phenomena are expected to manifest themselves most clearly for
 low momentum particles, e.g.\ as illustrated in
 \cite{bib:WWOXFORD}. Therefore, studies of the fragmentation
 function, i.e.\ the distribution of the scaled momentum,
 $\xp=p/\Ebeam$, are also relevant. The shape of a charged particle
 multiplicity distribution may be quantified by its dispersion
 (r.m.s.), D, and so \DQQQQ\ and \DQQLV\ are measured.

 Some earlier estimates of the sensitivity to colour reconnection have
 been made within the context of given models, comparing
 ``reconnection'' to ``no reconnection'' scenarios for \WWqqqq\
 events. In general, both the size and sign of any changes are
 dependent upon the model considered. At the expense of a reduction in
 statistical sensitivity, the dependence on the modelling of single
 hadronic W decays can be avoided by comparing directly the properties
 of the hadronic part of \WWqqln\ events with \WWqqqq\ events.  In the
 current study, the inclusive charged particle multiplicity and the
 fragmentation function are measured for \WWqqqq\ and the non-leptonic
 component of \WWqqln\ events.  Charged particles associated with the
 leptonically decaying W are excluded from these measurements.  The
 quantities $\Dnch=\nchQQQQ-2\nchQQLV$, $\DD=\DQQQQ-\sqrt{2}\DQQLV$
 and $\Dxp=\xpQQQQ-\xpQQLV$ are examined.  The mean values of the
 thrust distribution, \thrQQQQ, and the rapidity distribution relative
 to the thrust axis, \yQQQQ, are measured to characterise the global
 properties of \WWqqqq\ events themselves.  These are not expected to
 be particularly sensitive to the effects of colour reconnection
 \cite{bib:SK}.

 Models of colour reconnection as implemented in the event generators
 \Pythia\ (\SK\ model \cite{bib:SK}), \Ariadne\ (model \cite{bib:LL})
 and \Herwig\ (model \cite{bib:LEP2YR2,bib:HERWIG}) have been used in
 previous studies to assess the sensitivities of the quantities above
 \cite{bib:WWOXFORD}.  The models \Ariadne\ and a ``colour octet''
 variant\footnote{Merging of partons to form clusters was performed on
 a nearest neighbour basis, as a partial emulation of the model of
 Reference \cite{bib:EG1}.}  of \Herwig\ \cite{bib:BWOXFORD} predicted
 shifts in \nchQQQQ\ and \xpQQQQ\ similar in size to the expected
 statistical uncertainty on these quantities using the data studied in
 this letter.  However, in the studies of \cite{bib:WWOXFORD} no
 additional retuning of the models was performed after reconnection
 effects were included.
 
 In this letter, emphasis is placed on predictions from reconnection
 models which have been suitably tuned to describe \Zz\ data. The
 predictions from the reconnection models of \ARIADNE, \SK, and \EG\
 are considered, where the tuning of these models is summarised in
 Section~\ref{sec:CRmodels}. The version of \HERWIG\ including
 reconnection effects has not yet been tuned by OPAL and is therefore
 not discussed further herein.

\section{Data Selection}
\label{sec-selection}
 This analysis uses data recorded during 1997 by the OPAL detector,
 which is described fully elsewhere~\cite{bib:detector}.  The
 measurement of luminosity is identical to that in~\cite{bib:lumi183}.
 The integrated luminosity used in this analysis is
 $\intLdtfull\pm\dLstat\mathrm{(stat.)}
 \pm\dLsys\mathrm{(syst.)}$~pb$^{-1}$ at a luminosity weighted mean
 centre-of-mass energy of $\roots= \rroots \pm
 0.05$~\GeV~\cite{bib:energy183}.

 Events are selected in the \WWqqln\ and \WWqqqq\ channels using the
 likelihood selections described
 in~\cite{bib:opalmw172,bib:opalxs183}.  In total, 433 \WWqqqq\ and
 361 \WWqqln\ candidates are selected.  The Monte Carlo models without
 colour reconnection and the detector simulation are identical to
 those in~\cite{bib:opalxs183}.  The models of colour reconnection
 studied are described in Section~\ref{sec:CRmodels}.

 Charged particles may have up to 159 hits in the jet chamber. Tracks
 used in the analysis are required to have a minimum of 40 hits in the
 $|\cos\theta|$ region\footnote{The OPAL coordinate system is defined
 such that the origin is at the geometric centre of the jet chamber,
 $z$ is parallel to, and has positive sense along, the e$^-$ beam
 direction, $r$ is the coordinate normal to $z$, $\theta$ is the polar
 angle with respect to +$z$ and $\phi$ is the azimuthal angle around
 $z$.} in which at least 80 are possible.  At larger $|\cos\theta|$,
 the number of hits is required to be greater than 50\% of the
 expected number and also greater than 20.  Tracks must have a
 momentum component in the plane perpendicular to the beam axis of
 greater than 0.15~GeV$/c$, and a measured momentum of less than
 100~GeV$/c$. The extrapolated point of closest approach of each track
 to the interaction point is required to be less than 2~cm in the
 \mbox{$r$-$\phi$} plane and less than 25~cm in $z$.  Clusters of
 energy in the electromagnetic calorimeter were required to have a
 measured energy greater than 0.10~GeV (0.25~GeV) if they occurred in
 the barrel (endcap) region of the detector.

\section{Data Analysis and Correction Procedure}
\label{sec-analysis}

 The analysis of the event properties follows the unfolding procedure
 described in~\cite{bib:opalmw172}.  The distributions of \nch, \xp,
 $1-T$ and $y$ are corrected for background contamination using a
 bin-by-bin subtraction of the expected background, based on Monte
 Carlo estimates.  Corrections (described below) are then applied for
 finite acceptance and the effects of detector resolution, after which
 mean values of the distributions are calculated.  Each observable is
 evaluated using two samples of \eeWW\ events generated using the
 \KORALW~\cite{bib:KORALW} Monte Carlo program.  The first, which
 includes initial state radiation and a full simulation of the OPAL
 detector, contains only those events which pass the cuts applied to
 the data (detector level).  The second does not include initial state
 radiation or detector effects and allows all particles with lifetimes
 shorter than $3\times10^{-10}$~s to decay (hadron level).  Both
 samples are generated at the same $\sqrt{s}$.  Distributions
 normalised to the number of events at the detector and the hadron
 level are compared to derive bin-by-bin correction factors which are
 used to correct the observed distributions of \xp, $1-T$ and $y$ to
 the hadron level. In contrast to the other observables which are
 formed from charged particles alone, thrust is measured using both
 charged particles and clusters of electromagnetic energy unassociated
 with charged particles.

 A bin-by-bin correction procedure is suitable for the quantities
 above as the effects of finite resolution and acceptance do not cause
 significant migration (and therefore correlation) between bins.  Such
 a method is not readily applicable to multiplicity distributions, due
 to the large correlations between bins.  Instead, a matrix correction
 is used to correct for detector resolution effects, followed by a
 bin-by-bin correction which accounts for the residual effects due to
 acceptance cuts and initial state radiation, as in previous OPAL
 multiplicity studies \cite{bib:LEP1.5QCD}.

\begin{figure}[htbp]
 \centerline{\epsfig{file=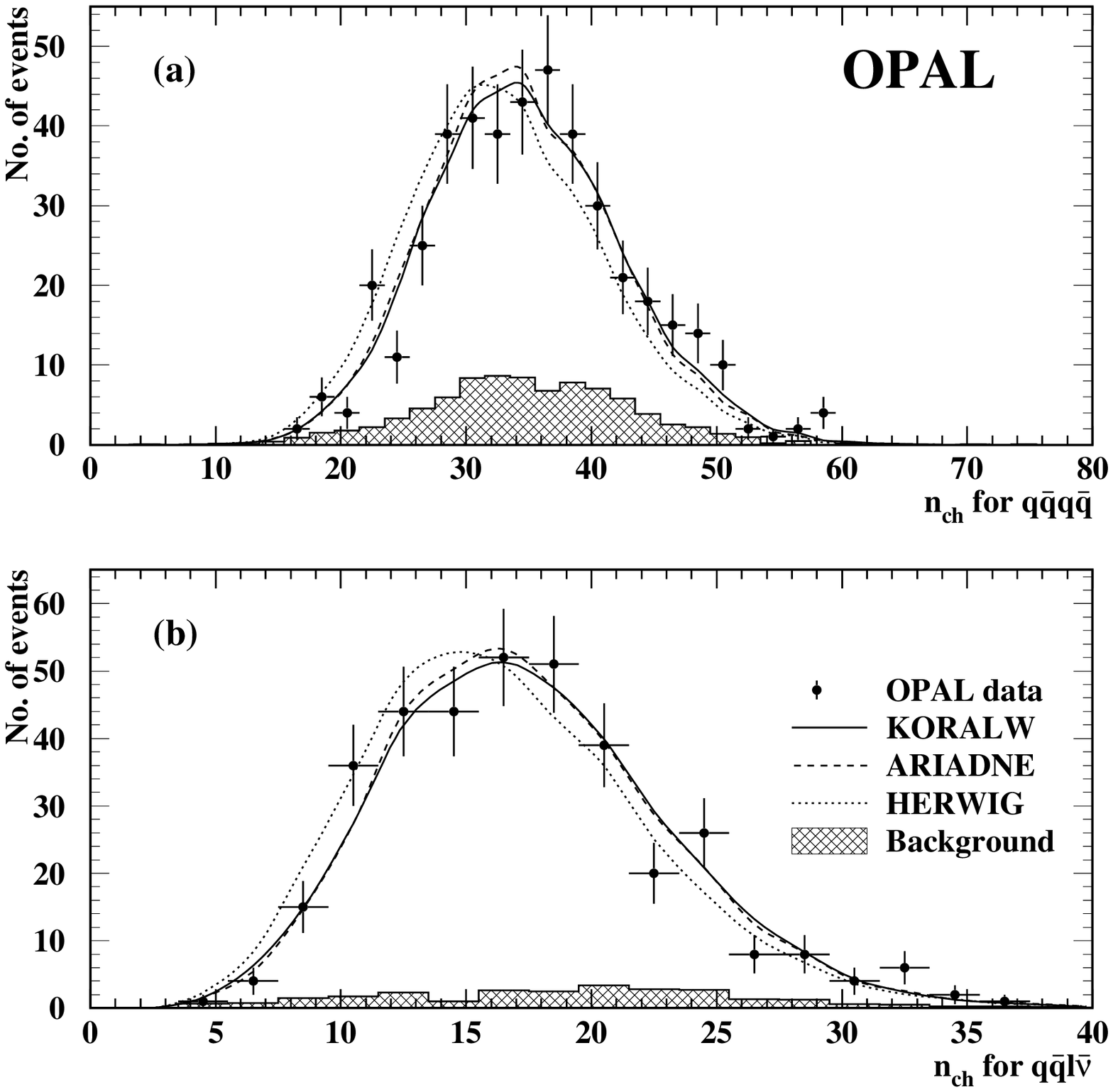,width=\textwidth}}
 \caption{Uncorrected charged multiplicity distributions for (a) \WWqqqq\
 events and (b) the hadronic part of \WWqqln\ events.  Points indicate the
 data, smooth curves show the expected sum of signal and background
 contributions for a variety of signal models, and the hatched histogram shows
 the expected background. Predictions of \KORALW, \PYTHIA\ and \EXCALIBUR\ are
 indistinguishable from one another.}  \label{fig-wwprop-nch}
\end{figure}

 The uncorrected multiplicity distributions for the \WW\ candidate
 events before background subtraction are illustrated in
 Figure~\ref{fig-wwprop-nch}, together with the predictions of Monte
 Carlo events without colour reconnection, but including detector
 simulation. The background prediction is the sum of all Standard
 Model processes, as described by the models used
 in~\cite{bib:opalmw183}. In general, good agreement is seen between
 the data and predictions from the models.  \HERWIG\ predicts visibly
 lower charged particle multiplicities than the other models, for
 example almost two units lower than that of \KORALW\ for \WWqqqq\
 events.  This is a consequence of the poor modelling of the
 multiplicity difference between quark flavours in the OPAL tuning of
 \HERWIG\ to \Zz\ data, and of the negligible b quark content of W
 decays.

\section{Colour Reconnection Models}
 \label{sec:CRmodels}

 Variants of three models of colour reconnection have been
 investigated.  The \SK\ (SK) models are based upon the Lund string
 picture of colour confinement, in which a string is created that
 spans the decay product partons associated with each W. These strings
 expand from the respective decay vertices and subsequently fragment
 to hadrons.  Before this occurs, at most one reconnection is allowed
 between sections of the two strings. The main scenarios considered
 are called type I and type II in analogy to the two types of
 superconducting vortices which could correspond to colour strings.
 In the \SKI\ scenario, the two colour flux tubes have a lateral
 extent comparable to hadronic dimensions. The probability for
 reconnection to occur is proportional to the integrated space-time
 volume over which the two tubes cross, where a (free) strength
 parameter, $\rho$, determines the absolute normalisation.  In the
 \SKII\ scenario, the strings have infinitesimally small radii and a
 unit reconnection probability upon their first crossing. A third
 scenario considered, \SKIIpr, is similar to \SKII\ but reconnection
 is only allowed to occur at the first string crossing which would
 reduce the total string length of the system.  As described in
 \cite{bib:SK98}, the only tuning necessary for these models is to
 ensure that the \JETSET\ hadronisation model gives a good description
 of \Zz\ data. Therefore, the same parameters were used as for the
 corresponding sample of non-reconnected \eeWW\ events, which for the
 \SK\ models are generated using \PYTHIA.  The fractions of \WWqqqq\
 events in which reconnection occurs at $\roots=183$~\GeV\ are found
 to be 37.9\% for \SKI\ (using\footnote{ $\rho=0.4$ results in a
 similar reconnection probability to the \SKII\ model.  The value used
 here increases the sensitivity of the data available by enhancing the
 effects of reconnection predicted by \SKI.} $\rho=0.9$), 22.1\% for
 \SKII\ and 19.8\% for \SKIIpr.

 The second set of two models is incorporated into the \ARIADNE\ Monte
 Carlo program by its author. They may be considered as extensions of
 the earlier model\footnote{In \cite{bib:GH}, at most one reconnection
 was allowed per event and possible reconnections within a single W
 were not implemented.} by \GH\ \cite{bib:GH}, as both models were
 implemented using the \ARIADNE\ Monte Carlo program and the same
 criterion is employed in the reconnection ansatz to determine whether
 reconnection is allowed.  Perturbative QCD favours configurations
 with minimal string length in hadronic \Zz\ decays
 \cite{bib:LEP2YR1}. When the partons of two W bosons are separating
 and strings formed between them, it is plausible that configurations
 corresponding to a reduced total string length are favoured. In the
 reconnection model of \ARIADNE, the string length is defined in terms
 of the $\Lambda$ measure, which may be viewed as the rapidity range
 along the string: $\Lambda=\sum_i\ln(m_i^2/m_\rho^2)$, where $m_i$ is
 the invariant mass of string segment $i$ and $m_\rho$ sets a typical
 hadronic mass scale. Reconnections are allowed, within constraints of
 colour algebra factors, which lead to a reduction in the total
 $\Lambda$ of the system. The first \ARIADNE\ model considered,
 referred to herein as \ARMII, restricts reconnections to gluons
 having energies less than \Gw, while the second \ARIADNE\ model,
 \ARMIII, does not impose such a restriction.  As gluons emitted with
 energies $>\Gw\sim 2$~GeV are perturbative in nature and have been
 shown to be radiated incoherently by the two initial colour dipoles
 \cite{bib:SK}, the model \ARMIII\ is disfavoured on theoretical
 grounds.  Multiple reconnections per event are permitted and
 reconnection may occur within different string segments of the same W
 boson.  The tuning of the models to describe \Zz\ data is as given in
 \cite{bib:ALEPH_tune}, with the following changes to the ``$a$''
 parameter, \texttt{MSTJ(41)}, which governs the hardness of the
 fragmentation function: non-reconnected \ARIADNE: $a=0.52$, \ARMII:
 $a=0.65$, \ARMIII: $a=0.58$.  The fractions of \WWqqqq\ events in
 which reconnection occurs at $\roots=183$~\GeV\ are found to be
 51.9\% for \ARMII\ and 63.4\% for \ARMIII.

 The third class of models considered is that of Ellis and Geiger
 \cite{bib:EG1}, as implemented in the \VNI\ Monte Carlo
 \cite{bib:VNI}, version 4.12. This model has been used to predict
 relatively large changes in observables such as the mean charged
 particle multiplicity \cite{bib:EG2}, but has so far not been
 subjected to significant comparison with \WW\ data. This model has
 three variants, called ``colour blind'', ``colour singlet'' and
 ``colourful''. In the colour blind scenario, colour degrees of
 freedom are ignored and cluster formation from partons proceeds
 solely on a nearest neighbour basis in space-time. The colour singlet
 case requires the colour degrees of freedom of two partons to add up
 to a colour singlet before cluster formation may take place.  In the
 colourful variant, partons which are not in a relative colour singlet
 state are also allowed to merge with each other, with the net colour
 degrees of freedom being balanced by the emission of additional,
 coloured partons. Predictions presented using this model correspond
 to the colour blind and colour singlet cases.  Predictions from the
 colourful case are not shown due to technical problems.  To generate
 \eeWW\ events, with the four fermions generated by \Pythia, it was
 necessary to change\footnote{\texttt{MSTW(3)=2}, \texttt{MSTV(84)=0,
 2} (blind, singlet), \texttt{MSTV(16)=1}, \texttt{MSTV(150)=2}.}  the
 default value of several program parameters.  The default tuning of
 the model is used, with only minor modifications\footnote{To ensure
 that unstable particles are decayed irrespective of their origin, and
 that charged particles carry a fraction of the event energy
 consistent with naive expectation assuming that pions dominate the
 final state, two changes are made: \texttt{MSTV(91)=5} and
 \texttt{MSTV(94)=0}.}.  In evaluating predictions from this model,
 decays of \Ks, $\Lambda^0$ and $\pi^0$ were included for consistency
 with the definition of stable particles given in
 Section~\ref{sec-analysis} (not the default for \VNI).  Significant
 energy imbalances were observed in hadronic \WW\ final states when
 comparing the generated particle spectra to the collision energy of
 183~GeV: approximately 3\% of \WWqqqq\ events differed from 183~GeV
 by more than 0.5~GeV, while approximately 3\% of \WWqqln\ events had
 deviations of more than 20~GeV.  These imbalances are unrelated to
 changes in the program parameters.

 Results obtained using the \EG\ model in \VNI\ are found to be rather
 different to those of \cite{bib:EG1,bib:EG2}.  Differences predicted
 between the charged multiplicities in the \WWqqqq\ and \WWqqln\
 channels are $\sim$2\% of \nchQQQQ, with opposite signs in the colour
 blind and colour singlet models.  It is noted that with the current
 version of \VNI, the predicted value of \thrQQQQ\ is found to be
 0.30. This is considerably larger than that of 0.06 given in
 \cite{bib:EG1}. A small value of \thrQQQQ\ indicates that an event
 has a two-jet like topology, allowing the hadronic showers from
 neighbouring W bosons to overlap, which in general is expected to
 enhance interconnection effects.

\begin{figure}[htbp]
 \centerline{\epsfig{file=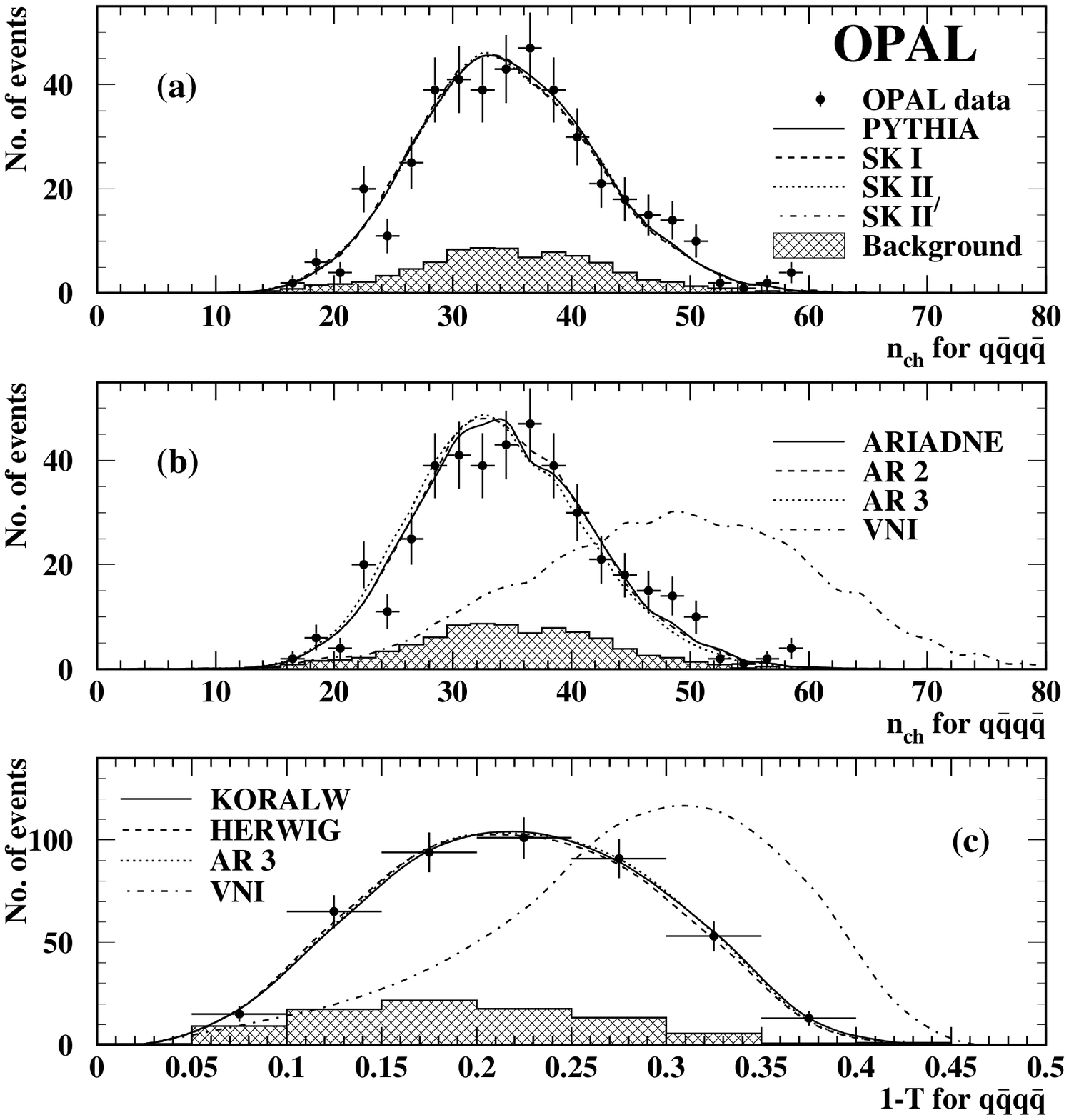,width=\textwidth}} \caption{Uncorrected
 charged multiplicity distributions for \WWqqqq\ events, compared with (a)
 \PYTHIA\ and the three \SK\ models of colour reconnection, and (b) \ARIADNE\
 and \EG\ (colour blind) models of colour reconnection.  Similarly, (c)
 compares the uncorrected thrust distribution for \WWqqqq\ events with a
 selection of the models, with and without colour reconnection.  Points
 indicate the data, smooth curves show the expected sum of signal and
 background contributions for a variety of signal models, and the hatched
 histogram shows the expected background.}  \label{fig-wwprop-nch-rec}
\end{figure}

 The \SK, \ARIADNE\ and \EG\ models, after simulation of the OPAL
 detector, are compared with the data distributions of charged
 particle multiplicity and thrust in the \WWqqqq\ channel.
 Figure~\ref{fig-wwprop-nch-rec}(a) presents the same data as in
 Figure~\ref{fig-wwprop-nch}(a) but compares them with the predictions
 of the \PYTHIA\ and \SK\ models, while
 Figure~\ref{fig-wwprop-nch-rec}(b) compares these data with the
 \ARIADNE\ and \EG\ (colour blind) models.
 Figure~\ref{fig-wwprop-nch-rec}(c) shows the thrust distribution with
 a subset of these models.  The model predictions are represented by
 smooth curves which pass through the centre of the binned predictions
 of each model.  It can be seen that the multiplicity predicted by the
 \EG\ model is significantly higher than observed in data or predicted
 by other models studied. The same observations are made when
 comparing this model to \WWqqln\ data.  The \ARMIII\ model predicts a
 charged multiplicity that is approximately 0.7 units lower than the
 non-reconnected \ARIADNE\ after detector simulation. From the $1-T$
 distribution, it is noted that most models provide a good description
 of data, while the \EG\ model predicts events which are much more
 spherical.  With its current implementation and tuning in the \VNI\
 program, the \EG\ model does not describe the data and is therefore
 not used to assess systematics on \Mw\ (see Section~\ref{sec-Mw}).

\section{Systematic Uncertainties}

 A number of systematic uncertainties have been considered in the
 analysis, as summarised in Tables~\ref{tab-wwnch-syst} and
 \ref{tab-wwprop-syst}. Each systematic uncertainty is taken as a
 symmetric error and the total uncertainty is defined by adding the
 individual contributions in quadrature.  Correlations are explicitly
 accounted for in estimating systematic uncertainties for \Dnch, \Dxp\
 and \DD.  The dependence of the correction procedure on the Monte
 Carlo model is evaluated by comparing results obtained using \KORALW,
 \PYTHIA\ and \EXCALIBUR~\cite{bib:EXCALIBUR} as the \WW\ signal
 samples, each using the same tuning of the \JETSET\ \cite{bib:PYTHIA}
 hadronisation model. The same treatment is applied to estimate model
 dependence in the subtraction of the small ( $< \cal O$ (1\%)) \WW\
 background contamination in each channel: \WWqqqq\ or \WWlnln\ events
 selected as \WWqqln, for example.  The \WW\ cross-section obtained
 from \GENTLE~\cite{bib:GENTLE} was allowed to vary from its central
 value of \GENTfull~pb by $\pm 4.3$\%, corresponding to the combined
 statistical and systematic uncertainty on the measured cross-section
 \cite{bib:opalxs183}.

 Hadronisation effects are estimated in a similar manner, unfolding
 data using \HERWIG\ and \ARIADNE,\footnote{The \ARIADNE\ events were
 generated using a different tune of \JETSET.} and also each of the
 variants of the \SK\ and \ARIADNE\ colour reconnection models.  The
 largest variation in any of these seven unfolding tests, dominated by
 the \ARMIII\ and \HERWIG\ models, was taken as the systematic
 uncertainty associated with hadronisation effects.  The \EG\ model
 was not used for unfolding as it does not describe the observed \WW\
 data.

 Uncertainties arising from the selection of charged tracks are
 estimated by varying the track selection cuts and repeating the
 analysis.  The maximum allowed values of the distances of closest
 approach to the interaction region in $r$-$\phi$ and $z$ are varied
 from 2~cm to 5~cm and from 25~cm to 50~cm, respectively, and the
 minimum number of hits on tracks is varied from 20 to 40. The
 dependence on charged track quality cuts is the sum in quadrature of
 these three effects.

 The dependence on the modelling of the accepted background is the sum
 in quadrature of two components, accounting separately for
 uncertainties in normalisation and shape.  Normalisation dependence
 is estimated by taking the largest effect observed when scaling
 various backgrounds before subtraction from the data.  The scale
 factors applied, taken from the uncertainties estimated in
 \cite{bib:opalxs183}, are approximately $\pm 10$\% for the total
 background in the \qq\qq\ channel and $\pm 8$\% in the \qq\lnu\
 channel, and approximately $\pm 11$\% in $\Zgamma\rightarrow\qq$
 background alone in both channels.  Two-photon backgrounds are scaled
 by a factor of 0.5 or 2, while the $\Zgamma\rightarrow\tau^+\tau^-$
 background is neglected.  Shape dependence of the subtracted
 distribution is estimated by taking the largest variation found when
 using alternative models for a given source of background, such as
 \HERWIG\ in place of \PYTHIA\ for the \Zgamma\ process, and \PYTHIA\
 in place of \PHOJET\ \cite{bib:PHOJET} for part of the two-photon
 process. In the case of the multiplicity distributions, the shape
 dependence was also tested by shifting the multiplicity by $\pm1$
 unit. This latter variation accommodates the differences between data
 and models, and the uncertainty on measured hadronic multiplicity in
 $\Zgamma\rightarrow\qq$ at $\roots\simeq182.7$~GeV
 \cite{bib:l3qcd183}.

  The four-fermion background was evaluated by comparing the accepted
 distributions from two samples of events generated using the
 \grcff~\cite{bib:GRC4F} model: one contains the full set of
 interfering four-fermion diagrams, while the other is restricted to
 the \CC\ set of W pair production diagrams\footnote{The W pair
 production diagrams, {\em i.e.}  $t$-channel $\nu_{\mathrm{e}}$
 exchange and $s$-channel \Zgamma\ exchange, are referred to as
 ``\CC'', following the notation of Ref.\ \cite{bib:LEP2YR2},
 p.~11.}. The same procedure was followed using \EXCALIBUR\ events.  A
 further test of the four-fermion contribution used independent
 samples of $\mathrm{W}\enu$, \ZZ\ and $\Zz\epem$ final states,
 generated with the \grcff\ or \PYTHIA\ Monte Carlo models.  The
 systematic uncertainty from the four-fermion background was taken to
 be the largest deviation seen from the results obtained using \grcff.

 Since most of the Monte Carlo samples used in the study were
 generated at $\roots=183$~GeV while the data were taken at various
 centre-of-mass energies between $\sim 182$ and 184~GeV, the analysis
 was repeated using \WW\ samples generated at 182~GeV and 184~GeV.
 The larger effect quantifies the dependence of the analysis on the
 choice of centre-of-mass energy.

 To estimate the sensitivity of the results to the unfolding
 procedure, the dispersions of the multiplicity distributions and the
 mean values \nchQQQQ, \nchQQLV, \xpQQLV, \xpQQQQ, \thrQQQQ\ and
 \yQQQQ\ were each evaluated by applying a single correction factor to
 each of the uncorrected values, rather than using the methods
 described in Section~\ref{sec-analysis}.  This correction is the
 ratio of the \KORALW\ prediction without detector simulation or
 initial state radiation, to the corresponding prediction for the same
 observable when these two effects are included.  The change in the
 corrected value gives an estimate of the systematic error due to the
 unfolding process.

 As a final test, the analysis was repeated using additional event
 selection criteria, based on the probability obtained from kinematic
 fits. These required energy and momentum conservation and constrained
 the masses of the two W boson candidates to be equal.  Consistent
 results were obtained and no additional systematic contributions were
 assigned.

\section{Results}
\label{sec:results}

The measurements of the mean charged particle multiplicities,
fragmentation functions, and their associated differences, corrected
to the hadron level, are:
\begin{eqnarray*}
           \nchQQQQ & = & 39.4 \pm 0.5 \pm 0.6 \\
           \nchQQLV & = & 19.3 \pm 0.3 \pm 0.3 \\
           \Dnch    & = & +0.7 \pm 0.8 \pm 0.6 \\
           \xpQQQQ  & = & (3.16  \pm 0.05  \pm 0.03)\times 10^{-2} \\
           \xpQQLV  & = & (3.25  \pm 0.07  \pm 0.04)\times 10^{-2} \\
           \Dxp     & = & (-0.09 \pm 0.09  \pm 0.05)\times 10^{-2}
\end{eqnarray*}
 where in each case the first uncertainty is statistical and the
 second systematic. Similarly corrected measurements of \yQQQQ,
 \thrQQQQ\ and the dispersions of the charged particle multiplicity
 distributions, are also presented in the upper part of
 Table~\ref{tab-mc-compare}.

  These results may be compared with the predictions from various
  Monte Carlo models, with and without reconnection effects, given in
  the lower part of Table~\ref{tab-mc-compare}. The predictions of the
  \EG\ model in \VNI\ are markedly different from data and other
  models.  The mean thrust indicates that the events are more
  spherical than data, the momentum spectrum is significantly softer
  than data, and the mean charged particle multiplicities are
  significantly higher than in data. Although \VNI\ predicts a much
  softer momentum spectrum and higher charged multiplicity than data,
  the behaviour is similar in both \qqqq\ and \qqln\ channels.

  Good agreement is found between the data and the standard QCD
  models, with the exception of \HERWIG\ as discussed in
  Section~\ref{sec-analysis}.  The higher value of \nchQQQQ\ measured
  in data than predicted by models can be inferred from the
  uncorrected multiplicity distribution of
  Figure~\ref{fig-wwprop-nch}(a), where a shift of approximately one
  unit relative to the corresponding predictions from \KORALW\ and
  \ARIADNE\ may be seen.  The \WWqqln\ data are in good agreement with
  these models.  This difference of approximately one unit in
  multiplicity is reflected in the final \Dnch\ obtained for the data.

 It is interesting to note that at the current precision, the mean
 charged particle multiplicity of a single hadronic W decay (\nchQQLV\
 or $\nchQQQQ/2$) is consistent with that of approximately 19.5 units
 predicted for $\Zgamma\rightarrow\qq$ at $\roots\simeq\Mw$, despite
 the different flavour composition expected in the two
 cases\footnote{The different flavour composition is predicted to
 lower the charged particle multiplicity of a single hadronic W decay
 by $\sim0.6$ units relative to $\Zgamma\rightarrow\qq$.}.  The
 predicted multiplicity is obtained from a fit \cite{bib:LEP1.5QCD} to
 the NLLA QCD calculation for the energy evolution of the charged
 particle multiplicity \cite{bib:nch_roots}, using data between 12~GeV
 and 133~GeV.

 The difference in mean charged multiplicities in hadronic W decays in
 \qq\qq\ and \qq\lnu\ events, \Dnch, is found to be consistent with
 zero at the current level of statistical precision. The dispersions
 in these two channels, which should be related by a factor $\sqrt{2}$
 in the case where two hadronically decaying W bosons in a given event
 are independent of each other, are also found to be consistent with
 each other.  Similarly, the measurements of the mean scaled charged
 particle momenta are consistent in the two channels. In contrast, the
 colour reconnection models predict non-zero values of \Dnch\ and
 \Dxp, as shown in Table~\ref{tab-mc-compare}.  The largest effects
 are seen for the \ARMIII\ model which is the least consistent with
 the measurements of \Dnch\ and \Dxp.

\begin{figure}[tp]
 \vspace*{-20mm}
 \centerline{\epsfig{file=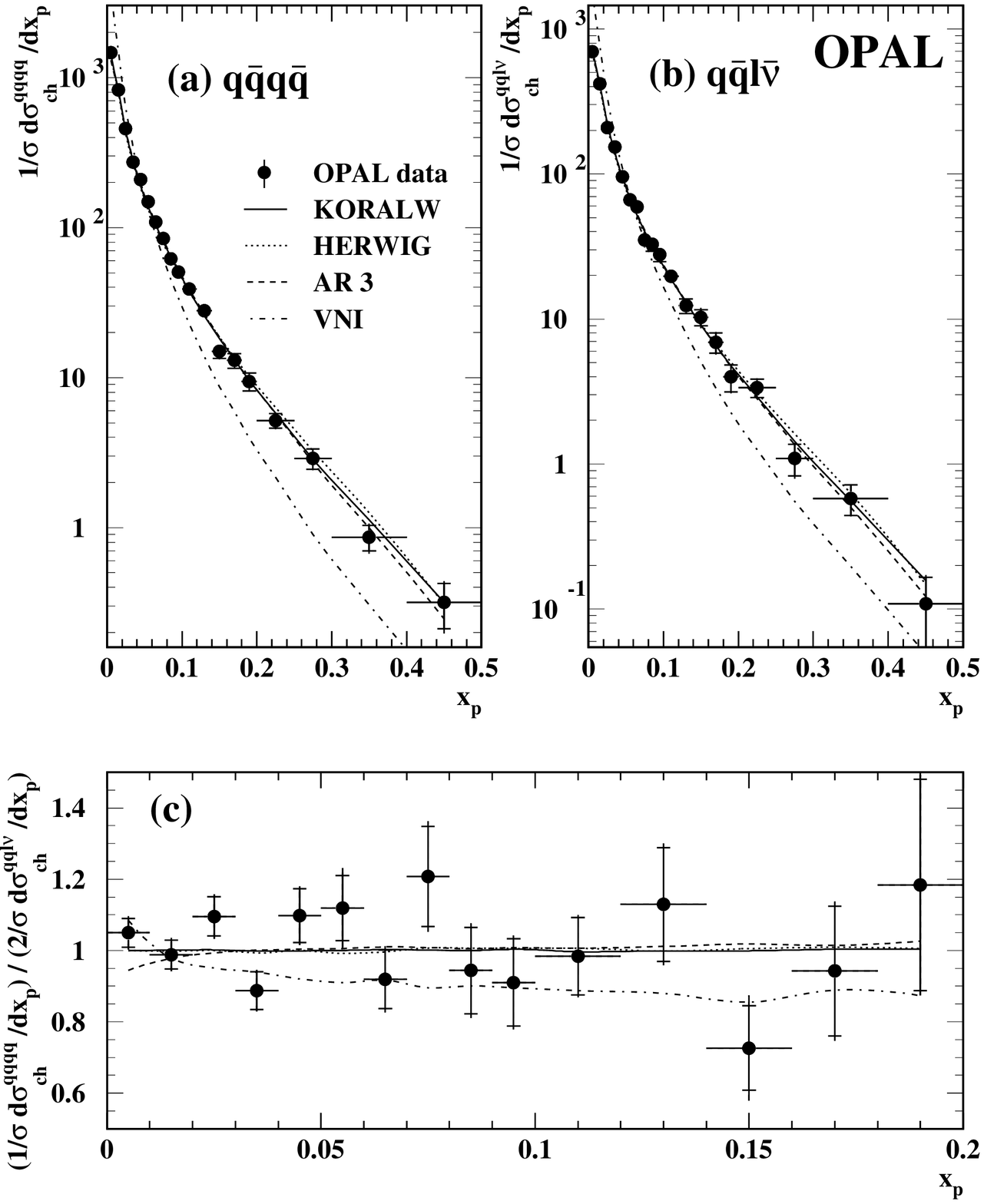,width=\textwidth}}
 \caption{Corrected \protect\xp\ distributions for (a) \WWqqqq\ events, (b)
 the hadronic part of \WWqqln\ events and (c) the ratio of the \WWqqqq\
 distribution to twice the \WWqqln\ distribution.  Points indicate the data,
 with statistical (horizontal bars) and systematic uncertainties added in
 quadrature. Point to point correlations exist in the systematic
 uncertainties.  Predictions of various Monte Carlo models (with and without
 colour reconnection) are shown as smooth curves.  \VNI\ predictions
 correspond to the colour singlet variant of the \EG\ model.  \SK\ and \ARMII\
 models are not shown as they are essentially indistinguishable from \KORALW.}
 \label{fig-wwprop-xp}
\end{figure}

  Figures~\ref{fig-wwprop-xp}(a) and (b) show the corrected
 fragmentation functions for the \WWqqqq\ and \WWqqln\ channels,
 together with predictions from the \KORALW, \HERWIG, \ARMIII\ and
 \VNI\ models.  Most models are in good agreement within statistical
 uncertainties in both cases. An alternative measurement of the mean
 charged multiplicity may be obtained from the integral of the
 fragmentation function.  The values determined in this way are
 $\nchQQQQ=39.5\pm0.9$ and $\nchQQLV=19.3\pm0.5$, where the errors
 given are combined statistical and systematic uncertainties.
 Figure~\ref{fig-wwprop-xp}(c) shows the ratio of the $x_{p}^{\QQQQ}$
 distribution to twice the $x_{p}^{\QQLV}$ distribution for low
 particle momenta, $\xp<0.2$.  There is no significant discrepancy
 between any of the viable reconnection models and data. Although the
 behaviour of the \ARMIII\ model indicates differences between the two
 channels in the low momentum region where the effects of colour
 reconnection are expected to be enhanced, there no indication that
 the data show such a tendency.

\begin{figure}[htbp]
 \centerline{\epsfig{file=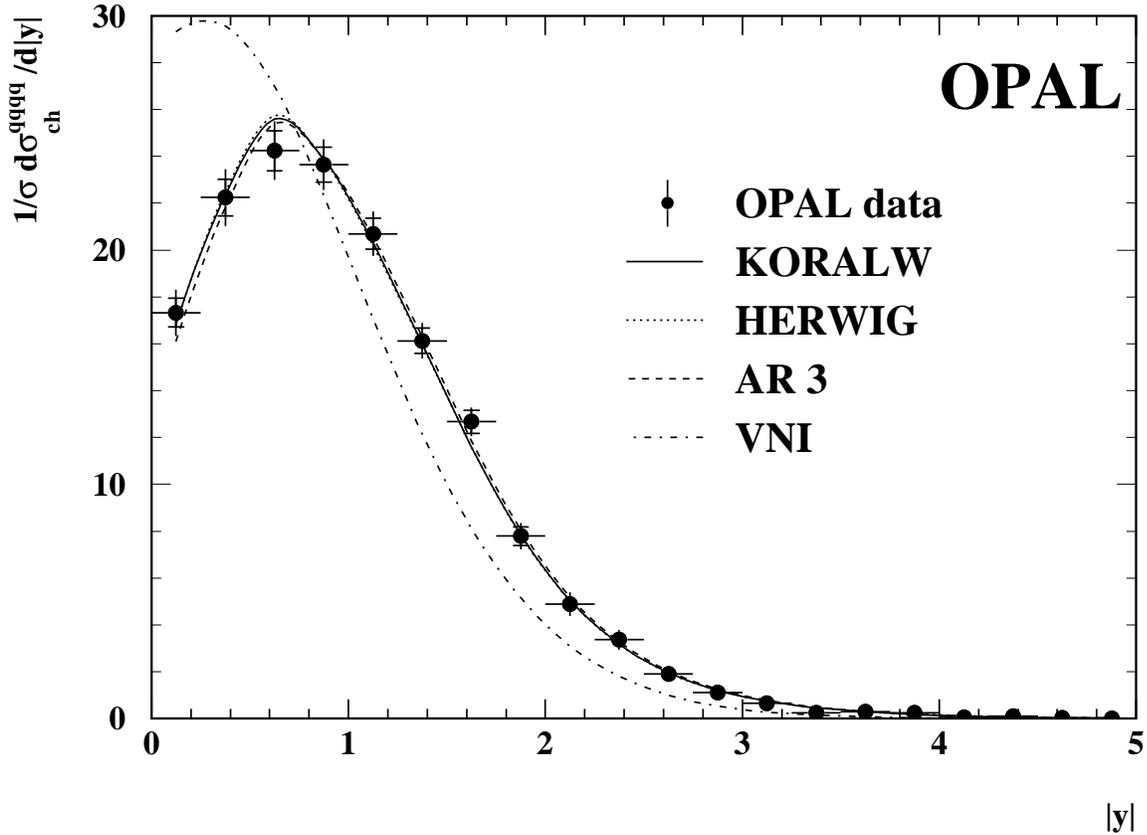,width=\textwidth}} \caption{Corrected
 rapidity distribution for \WWqqqq\ events. Points indicate the data, with
 statistical (horizontal bars) and systematic uncertainties added in
 quadrature.  Point to point correlations exist in the systematic
 uncertainties.  Predictions of various Monte Carlo models (with and without
 colour reconnection) are shown as smooth curves. \VNI\ predictions correspond
 to the colour singlet variant of the \EG\ model.  All models are normalised
 to the measured \nchQQQQ.  \SK\ and \ARMII\ models are not shown as they are
 essentially indistinguishable from \KORALW.}  \label{fig-wwprop-y}
\end{figure}

 The corrected rapidity distribution for \WWqqqq\ events, \y\, is
 shown in Figure~\ref{fig-wwprop-y}, together with the predictions of
 various Monte Carlo models. To focus attention on the shape of the
 distributions, the integrals of the model predictions have been
 normalised to the measured \nchQQQQ. All of the viable models provide
 a good description of the data over the entire range measured.

\section{Estimated effect on \Mw\ measurement}
\label{sec-Mw}
 The effect predicted by each model of colour reconnection on the OPAL
 measurement of \Mw\ in the \WWqqqq\ channel was investigated by
 analysing Monte Carlo \WWqqqq\ events, with simulation of the OPAL
 detector, as if they were data. In each selected event, three
 kinematic fits are performed corresponding to the possible jet-jet
 pairings, and further criteria are used to select up to two
 reconstructed masses per event, \Mrec, as described in the
 accompanying paper \cite{bib:opalmw183}.

 For each model a possible shift in the measured mass was estimated in
 three ways.  In the first method, a distribution of the
 event-by-event difference between \Mrec\ and \Mtrue\ was formed,
 where \Mtrue\ is defined as the average of the generated \Wp\ and
 \Wm\ masses.  Following the Breit-Wigner fitting analysis of
 \cite{bib:opalmw183}, up to two values of $(\Mrec-\Mtrue)$ may enter
 the distribution, depending upon their relative and absolute
 kinematic fit probabilities. Events admitted to the distribution are
 required to have radiated less than 0.1~GeV of energy into initial
 state photons.  The shift is estimated from the truncated mean, using
 a range of $\pm3$~GeV.  The results of this procedure are given in
 Table~\ref{tab-mw-bias} in the column ``Method 1a''.  A variant of
 this method, which accounts for possible biases in the mass
 reconstruction in the non-reconnected samples, defines the colour
 reconnection effect as the difference between the means in the
 corresponding reconnected and non-reconnected samples.  The results
 are given in column ``Method 1b''.

 In the second method, the reweighting analysis used to determine the
 principal result of \cite{bib:opalmw183} was applied to each sample
 of simulated events, giving a fitted mass, \Mfit.  Assuming that this
 procedure does not introduce any significant bias, the difference
 between \Mfit\ and the input \Mw\ value in the models is taken to be
 the change in measured \Mw\ due to colour reconnection.  The results
 of this procedure are given in Table~\ref{tab-mw-bias} in column
 ``Method 2a''.  The column ``Method 2b'' presents analogous results
 to Method 1b.

 It can be seen that the changes predicted for the non-reconnection
 models are small and consistent with zero. In contrast, a significant
 difference is visible for the \ARIADNE\ models, while effects for the
 \SK\ models are less pronounced. The largest change in \Mw\ is seen
 in the model \ARMIII\ which also predicts the most significant change
 in \Dnch.  This model is disfavoured theoretically as noted in
 Section~\ref{sec:CRmodels}, and experimentally by measurements
 presented in this letter.  Therefore, this model is not used to
 estimate possible systematic effects on \Mw\ measurements.

 The individual biases of the admissible models are used to estimate
 the influence which colour reconnection may have on measurements of
 \Mw.  Consistent results are obtained in the three methods.  Although
 Method 1b is the more statistically precise procedure, the overall
 systematic uncertainty is taken from Method 2b, which accounts
 explicitly for possible biases inherent in the actual procedure with
 which \Mw\ is extracted from data. The largest individual effect is
 seen from the \ARMII\ model, with a value of 49~MeV. This is
 therefore assigned as the current estimate of the systematic
 uncertainty associated with colour reconnection effects on OPAL
 measurements of \Mw.

\section{Conclusions}
\label{sec:conclusions}
 The \WW\ event properties presented here are in good agreement with
 expectations of standard QCD models.  A 10\% reduction in \nchQQQQ\
 as predicted by Ellis and Geiger \cite{bib:EG2} is not supported by
 data.  This difference in the charged multiplicity is not reproduced
 by the Monte Carlo program \VNI\ in which the \EG\ model is
 implemented.  The current implementation of the \EG\ model is
 excluded by the measured event shape data and therefore is not used
 in estimating the systematic uncertainty on \Mw\ from colour
 reconnection.

  Studies of reconnection phenomena implemented with the \Ariadne\ and
 \Pythia\ models show that changes up to approximately 3\% may be
 expected in \nchQQQQ\ and \xpQQQQ.  Defining \Dnch\ and \Dxp\ using
 data alone provides a model independent test of possible reconnection
 effects.  The maximum shifts in these variables predicted by the
 models considered are at the level 1--2 standard deviations for the
 current data.  There is no indication of the effects of colour
 reconnection on these observables at the current level of statistical
 precision.

 The effect of colour reconnection on measurements of \Mw\ is
 estimated using the \SK\ and \ARIADNE\ Monte Carlo models of colour
 reconnection which are theoretically admissible and consistent with
 the currently available data.  A model dependent estimate of
 $\Delta\Mw=\pm49$~MeV is currently assigned to the possible influence
 that colour reconnection may have on \Mw\ measurements in the
 \WWqqqq\ channel.

\medskip
\bigskip\bigskip
\appendix
\par
\section*{Acknowledgements}
We thank Klaus Geiger for his help and support with the computer program
\VNI, and note with sorrow his tragic death on 2 September 1998.

\noindent
We particularly wish to thank the SL Division for the efficient operation
of the LEP accelerator at all energies
 and for their continuing close cooperation with
our experimental group.  We thank our colleagues from CEA, DAPNIA/SPP,
CE-Saclay for their efforts over the years on the time-of-flight and trigger
systems which we continue to use.  In addition to the support staff at our own
institutions we are pleased to acknowledge the  \\
Department of Energy, USA, \\
National Science Foundation, USA, \\
Particle Physics and Astronomy Research Council, UK, \\
Natural Sciences and Engineering Research Council, Canada, \\
Israel Science Foundation, administered by the Israel
Academy of Science and Humanities, \\
Minerva Gesellschaft, \\
Benoziyo Center for High Energy Physics,\\
Japanese Ministry of Education, Science and Culture (the
Monbusho) and a grant under the Monbusho International
Science Research Program,\\
Japanese Society for the Promotion of Science (JSPS),\\
German Israeli Bi-national Science Foundation (GIF), \\
Bundesministerium f\"ur Bildung, Wissenschaft,
Forschung und Technologie, Germany, \\
National Research Council of Canada, \\
Research Corporation, USA,\\
Hungarian Foundation for Scientific Research, OTKA T-016660, 
T023793 and OTKA F-023259.

\clearpage

\clearpage


\begin{table}[htbp]
 \begin{center}
 \begin{tabular}{|l|c|c|c|c|c|c|} \hline
 Systematic variation & 
 \nchQQQQ & \nchQQLV & \Dnch & \DQQQQ   & \DQQLV & \DD \\
\hline
 \WW\ model            & $0.14$ & $0.03$ & $0.13$ & $0.13$ & $0.08$ & $0.16$ \\
 Hadronisation model   & $0.45$ & $0.21$ & $0.41$ & $0.45$ & $0.31$ & $0.11$ \\
 Track quality cuts    & $0.25$ & $0.12$ & $0.25$ & $0.15$ & $0.11$ & $0.08$ \\
 Background model      & $0.23$ & $0.05$ & $0.17$ & $0.03$ & $0.02$ & $0.02$ \\
 Four-fermion                                                    
          background   & $0.18$ & $0.05$ & $0.08$ & $0.07$ & $0.02$ & $0.06$ \\
 Centre-of-mass energy & $0.01$ & $0.11$ & $0.21$ & $0.06$ & $0.06$ & $0.08$ \\
 Unfolding procedure   & $0.12$ & $0.03$ & $0.17$ & $0.07$ & $0.06$ & $0.10$ \\
\hline                                                           
 Total                 & $0.62$ & $0.28$ & $0.60$ & $0.51$ & $0.35$ & $0.25$ \\
\hline
 \end{tabular}
 \end{center}
\caption{Individual contributions to the systematic uncertainties on charged
         multiplicity related quantities.}
\label{tab-wwnch-syst}
\end{table}

\begin{table}[htbp]
 \begin{center}
 \begin{tabular}{|l|c|c|c|c|c|c|c|c|} \hline
 Systematic variation    & \xpQQQQ   & \xpQQLV & \Dxp   & \thrQQQQ & \yQQQQ \\ 
                         & $\times 10^{2}$
                                     & $\times 10^{2}$ & $\times 10^{2}$ 
                                     & $\times 10^{2}$ & $\times 10^{2}$    \\ 
\hline
 \WW\ model              &   $0.012$ & $0.008$ & $0.018$ & $0.11$ & $0.28$  \\
 Hadronisation model     &   $0.012$ & $0.015$ & $0.019$ & $0.29$ & $0.81$  \\
 Track quality cuts      &   $0.019$ & $0.018$ & $0.015$ & $0.11$ & $0.31$  \\
 Background model        &   $0.003$ & $0.006$ & $0.005$ & $0.27$ & $0.96$  \\
 Four-fermion background &  $0.011$ & $0.019$ & $0.027$ & $0.08$ & $0.23$   \\
 Centre-of-mass energy   &   $0.012$ & $0.019$ & $0.018$ & $0.36$ & $0.38$  \\
 Unfolding procedure     &   $0.002$ & $0.019$ & $0.022$ & $0.68$ & $0.17$  \\
\hline                                                
 Total                   &   $0.031$ & $0.042$ & $0.049$ & $0.87$ & $1.4$   \\
\hline
 \end{tabular}
 \end{center}
\caption{Individual contributions to the systematic uncertainties on the
         average event properties.}
\label{tab-wwprop-syst}
\end{table}


\begin{table}[htbp]
\begin{sideways}%
\begin{minipage}[b]{\textheight}
 \begin{center}
 \begin{tabular}{|l|c||c|c|c||c|c|c||c|c|c||c|c|} \hline
 &  \Prec\ (\%)
 & \nchQQQQ & \nchQQLV & \Dnch & \DQQQQ   & \DQQLV & \DD &
  \xpQQQQ & \xpQQLV  & \Dxp  & \thrQQQQ & \yQQQQ  \\ 
 Sample   &          &       &          &        &     &
 & &  $\times 10^{2}$
          & $\times 10^{2}$
                     & $\times 10^{2}$
                                        &        &     \\
\hline   
Data     &
         & 39.4  & 19.3  & $+0.7$ & 8.8 & 6.1 & $+0.2$ &
            3.16 &  3.25 & $-0.09$ & 0.240 & 1.011 \\
stat. &   & $\pm0.5$ & $\pm0.3$& $\pm0.8$ &
           $\pm0.3$ & $\pm0.3$ & $\pm0.5$ &
           $\pm0.05$& $\pm0.07$  & $\pm0.09$ &
           $\pm0.015$ & $\pm0.014$ \\
syst. &   & $\pm0.6$ & $\pm0.3$& $\pm0.6$ &
           $\pm0.5$ & $\pm0.4$ & $\pm0.3$ & 
           $\pm0.03$& $\pm0.04$ & $\pm0.05$ &
           $\pm0.009$ & $\pm0.014$ \\

\hline                                                

\KORALW  & 0.0
         & 38.78 & 19.38 & $+0.01$ & 8.47  & 6.00 & $-0.01$ &
            3.21 &  3.21 & 0.00 & 0.240 & 1.009  \\
\HERWIG  & 0.0
          & 37.24 & 18.63 & $-0.02$ & 8.65  & 6.14 & $-0.03$ &
            3.31 &  3.30 & 0.01 & 0.239 & 1.009 \\

\PYTHIA  & 0.0
         & 38.77 & 19.39 & $-0.01$ & 8.49  & 5.97 &  $+0.05$   &
            3.21 &  3.21 & 0.00 & 0.240 & 1.009 \\

\SKI     & 37.9
         & 38.39 &  $-$  & $-0.38$ & 8.45  & $-$  & $+0.01$ &
            3.25 &  $-$  & $+0.04$ & 0.239 & 1.015 \\

\SKII    & 22.2
         & 38.59 & $-$   & $-0.19$ & 8.45  & $-$  & $+0.01$ &
            3.23 & $-$   & $+0.02$ & 0.240 & 1.010 \\

\SKIIpr  & 19.8
         & 38.54 & $-$   & $-0.24$ & 8.41  & $-$  & $-0.03$ &
            3.23 & $-$   & $+0.02$ & 0.240 & 1.011 \\

\ARIADNE & 0.0
         & 38.47 & 19.22 & $+0.04$ & 8.23  & 5.82 & 0.00    &
            3.22 &  3.22 & 0.00 & 0.240 & 1.009 \\

\ARMII   & 51.9
         & 38.31 & 19.31 & $-0.31$ & 8.01  & 5.67 & $-0.01$ &
            3.24 & 3.21  & $+0.03$ & 0.239 & 1.013 \\

\ARMIII  & 63.4
         & 37.45 & 19.17 & $-0.90$ & 7.96  & 5.67 & $-0.05$ &
            3.31 &  3.23 & $+0.08$ & 0.238 & 1.024 \\

\VNI (blind)    & $>0$
         & 68.5  & 35.0  & $-1.6$ & 14.8  & 10.6 & $-0.1$ &
            1.80 &  1.74 & $-0.06$& 0.30  & 0.81  \\

\VNI (singlet)  & $>0$
         & 70.3  & 34.4  & $+1.4$ & 15.5  & 10.7 & $+0.5$ &
            1.76 &  1.89 & $-0.13$& 0.31  & 0.80  \\
\hline
 \end{tabular}
 \end{center}
\caption{Summary of measurements performed, after correction to the hadron
level. The predictions from a variety of Monte Carlo models, with and without
colour reconnection effects, are also given.  The fraction of events in which
at least one reconnection has occurred in \WWqqqq\ events is indicated by the
column, \Prec.  Statistical uncertainties in the Monte Carlo predictions are
typically 1--2 units in the least significant digit. In the case of the \SK\
models which do not include reconnection for \WWqqln\ events, the predictions
from the corresponding ``no reconnection'' sample (\PYTHIA) were used to
determine the quantities \Dnch, \DD\ and \Dxp.}
\label{tab-mc-compare}
\end{minipage}\end{sideways}
\end{table}

\begin{table}[htbp]
 \begin{center}
 \begin{tabular}{|l|c|c|c|c|} \hline
  & \multicolumn{4}{|c|}{\Mw\ bias (MeV)} \\
 Model   &  Method 1a  & Method 1b   & Method 2a   & Method 2b \\ 
\hline   

\KORALW  & $-11\pm11$  &  $-$        & $-12\pm18$  & $-$ \\
		                     
\HERWIG  &  $-13\pm9$  &  $-$        & $-26\pm15$  & $-$ \\
		                     
\PYTHIA  &   $+2\pm9$  &  $-$        & $+30\pm18$  & $-$  \\
		                     
\SKI     &  $+43\pm9$  &  $+41\pm13$ & $+41\pm18$  & $+11\pm25$ \\
		                     
\SKII    &  $+1\pm9$   &  $-1\pm13$  & $+7\pm18$   & $-23\pm25$\\
		                     
\SKIIpr  &  $+4\pm9$   &  $+2\pm13$  & $+8\pm18$   & $-22\pm25$ \\
		                     
\ARIADNE &  $-10\pm5$  &  $-$        & $-7\pm10$   & $-$  \\
		                     
\ARMII   &  $+37\pm5$  &  $+47\pm7$  & $+42\pm10$  & $+49\pm14$ \\
		                     
\ARMIII  & $+100\pm10$ & $+110\pm11$ & $+138\pm18$ & $+145\pm21$ \\

\hline
 \end{tabular}
 \end{center}
\caption{Predictions for the bias in the measured \Mw\ from a variety of Monte
Carlo models, with and without colour reconnection effects.  Simulation of the
OPAL detector is included.}
\label{tab-mw-bias}
\end{table}

\clearpage

\end{document}